\documentclass[twocolumn,showpacs,preprintnumbers,amsmath,amssymb]{revtex4}
\usepackage{tabularx,graphicx}\begin{document}
\newcommand{\beq}{\begin{equation}}
\newcommand{\eeq}{\end{equation}}
\newcommand{\beqn}{\begin{eqnarray}}
\newcommand{\eeqn}{\end{eqnarray}}
\newcommand{\bmath}{\begin{subequations}}
\newcommand{\emath}{\end{subequations}}
\def\skipline{\par \vspace{\baselineskip}}
\title{The London moment: what a rotating superconductor reveals about superconductivity
}
\author{J. E. Hirsch }
\address{Department of Physics, University of California, San Diego\\
La Jolla, CA 92093-0319}
 
\begin{abstract} 
The London moment is the magnetic moment acquired by a rotating superconductor. We propose  that the London moment reveals
the following  fundamental properties of the superconducting state:
(i) superconductors (unlike normal metals) know the $sign$ of the charge carriers, (ii) the superconducting charge carriers are $free$ electrons, 
(iii) electrons are expelled from the interior to the surface in the transition to the superconducting state, (iv), superfluid electrons occupy orbits of
radius $2\lambda_L$ ($\lambda_L=$London penetration depth), and 
  (v) a spin current exists in the ground state of superconductors. These properties are consistent with the Meissner effect, however the
Meissner effect does not $directly$ reveal the sign of the charge carriers nor the fact that the carrier's mass is the free electron mass nor the fact
that a spin current exists in superconductors. Note also that within the BCS theory of superconductivity none of the key
properties of superconductors listed above are predicted. Instead, these properties are predicted by the theory of hole superconductivity.
\end{abstract}
\maketitle

\section{introduction}
While every book on superconductivity features prominently the Meissner effect, almost none, particularly the most popular ones, discuss the London moment.
For example, the books by  de Gennes\cite{degennes}, Schrieffer\cite{schrieffer}, Ketterson and Song\cite{ketterson}, Buckel\cite{buckel}, Tilley\cite{tilley} 
Abrikosov\cite{abrikosov} and Parks\cite{parks} don't even mention it, Tinkham's\cite{tinkham}, only in passing.  As a consequence, the phenomenon appears to be unknown to a large fraction of contemporary physicists working
in the field of superconductivity.

The London moment is the magnetic moment exhibited by a rotating superconductor. The phenomenon was predicted in 1933, just before the Meissner effect was discovered, 
in a seminal paper 
by Becker, Heller and Sauter\cite{becker}: a superconducting body  rotating with angular velocity $\vec{\omega}_0$ develops a uniform magnetic field throughout its interior, given by
\beq
\vec{B}=-\frac{2m_ e c}{e}\vec{\omega}_0
\eeq
with $e$ ($<0$) the electron charge, and $m_e$ the $bare$ electron mass. The magnetic field is $parallel$ to the angular velocity. The resulting magnetic moment will depend on the shape of
the body. For example for a sphere of radius $R$  the resulting  magnetic moment $\vec{m}$ is,
\beq
\vec{m}=-\frac{m_e c}{e}R^3\vec{\omega}_0
\eeq
 These results are valid for bodies of dimensions much larger than the London penetration depth. Qualitatively we can understand the physics as follows: when the superconductor is set into
 rotation, the electrons initially stay at rest. The electric current produced by the moving ions generates a changing magnetic field, which in turn generates an azimuthal electric field that
 sets the electrons into motion. Because of the inertia of the electrons they slightly `lag behind' the motion of the body near the surface, giving rise to an electric current that
 generates the magnetic field in the interior of the superconductor.
 
 The reason the phenomenon is called the London moment rather than the `Becker et al moment' is presumably that Becker et al predicted this phenomenon only for the case when a body at rest that is
 already superconducting is set into rotation. Instead, F. London realized,   $after$ the Meissner effect had been discovered experimentally, that a rotating normal metal
 cooled into the superconducting state while rotating would develop {\it the same} magnetic moment. In other words, just like for the Meissner effect, the state of a rotating superconductor is
 independent of its history.
 
 It is interesting to read about a key aspect of the phenomenon in London's own words\cite{londonbook}:
 \newline
 ``There is an implication which might be worth mentioning since it would appear quite absurd from the viewpoint of the perfect conductor concept. The uniqueness properties of the present
 theory provide for only $one$ current distribution {\it independent of the way} in which the superconducting state is reached. Consequently we have to conclude that the {\it same state} as has been
 calculated above must also be obtained if the sphere is cooled from the normal into the superconducting state while it is rotating. The perfect conductor theory would, of course,
 furnish no reason for the electrons near the surface of the metal to lag suddenly behind when the rotating sphere goes into the superconducting state. 
 It would simply lead to a state of zero magnetic moment in which all charges move in phase - the same below as above the transition point.''
 
 This bold prediction of London (in his words, ``there cannot be much doubt as to the outcome of this experiment''\cite{londonbook}) was first verified experimentally in 1964 by Hildebrandt
  for Pb\cite{lm1,lm2}. Since then it has been tested for other `conventional' superconductors\cite{lm3,lm4,lm5,lm6,lm7} as well as for `unconventional' ones including high $T_c$ cuprates\cite{lm8} and heavy fermions\cite{lm9}, yielding always results   in agreement with Eq. (1), independent
  of history.
  
  In this paper we propose that the London moment reveals fundamental properties  of the superconducting state that have been unrecognized in the conventional understanding of
  superconductors. 
  
    \begin{figure}
\resizebox{6.5cm}{!}{\includegraphics[width=7cm]{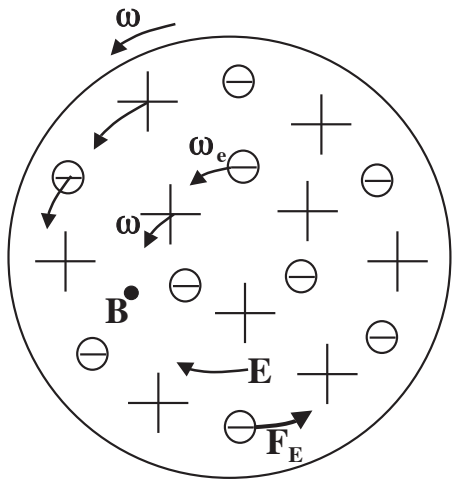}}
  \caption{The body rotates with an increasing angular velocity $\bold{\omega}$ in counterclockwise direction. The motion of the ions (positive charge) generates
  a magnetic field $\bold{B}$ pointing out of the paper that increases with time, inducing a Faraday electric field $\bold{E}$ in clockwise direction
  that exerts a counterclockwise force $\bold{F_E}$ on the electrons, driving them to follow the ionic rotation. The electrons in the interior rotate
  with the same angular velocity as the ions, and near the surface with slightly smaller angular velocity, as shown schematically
  by the arrows  }
\end{figure} 
  
  \section{classical derivation of the London moment}
  We start by reviewing the classical derivation of the effect as discussed by Becker et al\cite{becker} before the Meissner effect was discovered.
  
  Consider a perfect conductor at rest in the absence of external fields that is put into rotational motion starting at time $t=0$. The body has angular
  velocity $\omega$ at time $t$ and attains a final angular velocity $\omega_0$, it does not matter
  how long it takes to reach the final state. We assume, following Becker et al\cite{becker}, that the superfluid electrons are completely detached from the ions. As the body starts to rotate 
  only the positive ions rotate, giving rise to an electric current and a magnetic field. The changing magnetic field generates an electric field through Faraday's law, and this electric field acts on
  the superfluid electrons pushing them along to follow the ions. The direction of the fields and rotation  is shown in Fig. 1.
  
  Assume at time t an electric field E is acting on the electrons, exerting an azimuthal force
  \beq
  m_e \frac{dv}{dt}=eE
  \eeq
  with $v$ the electron's azimuthal speed. The body is rotating rigidly with angular velocity $\omega$, and we assume the electrons follow suit with angular velocity $w_e$ which
  is very close to $\omega$. In fact, we assume that $\omega_e$ is identical to $\omega$ in the bulk of the system, so that no volume currents exist, and only differs from $\omega$ near the
  surface giving rise to a surface electric current. We will see that this assumption is self-consistent. We have then $v=\omega_e r=\omega r$ for electrons in the bulk at distance $r$ from the rotation axis, hence from Eq. (1)
  \beq
  E=\frac{m_e}{e}\dot{\omega} r  .
  \eeq
 
 This electric field is generated by the changing magnetic field which in turn results from the ionic rotation giving rise to an electric current. From Faraday's law,
 \beq
 \oint \vec{E}\cdot \vec{dl}=-\frac{1}{c}\frac{\partial}{\partial t} \int \vec{B}\cdot\vec{dS}
 \eeq
 yielding at distance r
 \beq
 E=-\frac{1}{2c}r\dot{B}       .
 \eeq
Replacing $E$ from Eq. (4) into Eq. (6) and solving for $\dot{B}$
\beq
\dot{B}=-\frac{2m_e c}{e}\dot{w}
\eeq
and finally, integrating and using that $\omega(t=0)=B(t=0)=0$
\beq
B=-\frac{2m_e c }{e} \omega_0
\eeq
in agreement with Eq. (1).

This remarkably simple  derivation says remarkably little about the actual origin of the magnetic field, i.e. the nature and magnitude of the surface current
created by the lagging electrons. To obtain the current density
\beq
\vec{j}(\vec{r},t)=en_s[\vec{v}(\vec{r},t)-\vec{v}_0(\vec{r}))
\eeq
with $\vec{v}_0=\vec{\omega}_0\times\vec{r}$  the ionic speed one uses the equation of motion (3) together
with Maxwell's equations
\bmath
\beq
\vec{\nabla}\times\vec{B}=\frac{4\pi}{c}\vec{j}
\eeq
\beq
\vec{\nabla}\times\vec{E}=-\frac{1}{c} \frac{\partial \vec{B}}{\partial t}
\eeq
\emath
 and the boundary conditions that $\vec{B}$ and its derivatives are continuous across the surface of the body. 
Becker et al performed this calculation for a rotating sphere of radius $R$, and obtained the result for the superfluid velocity field
(in what follows $r$ denotes distance to the center of the sphere, not to the rotation axis):
\bmath
\beq
\vec{v}=\vec{\omega}_0\times \vec{r} [1+3\frac{\lambda_L^2R}{r^3}f(r)]
\eeq
\beq
f(r)= \frac{1}{sinh(R/\lambda_L)}[sinh(r/\lambda_L)-\frac{r}{\lambda_L} cosh(r/\lambda_L]
\eeq
\emath
from which the current Eq. (9) follows. Note that $f(r)$ is very small except for $r$ very near the surface. For $r>>\lambda_L$ it is given approximately by
\beq
f(r)=-\frac{r}{\lambda_L} e^{-(R-r)/\lambda_L}
\eeq
so that indeed $\vec{v}=\vec{\omega}_0\times\vec{r}$ for $R-r>>\lambda_L$, as assumed in the earlier derivation. In other words,
the superfluid velocity equals the ion velocity except close to the surface, so there is no volume current. Near the surface,
Eq. (11) becomes approximately
\beq
\vec{v}=\vec{\omega}_0\times\vec{r}[1-3\frac{\lambda_LR}{r^2}e^{-(R-r)/\lambda_L}]
\eeq
so that even at the surface the superfluid velocity is only slightly less than the body's velocity, yet sufficiently different to give rise
to the interior magnetic field Eq. (1). The full expression for the magnetic field, using spherical coordinates $(r, \theta, \varphi)$ with
$\vec{\omega}_0$ parallel to the $z$-axis  is\cite{becker}
\beq
\vec{B}=-\frac{2m_e c}{e}\vec{\omega_0}+\delta B_r\hat{r}+\delta B_\theta \hat{\theta}
\eeq
with
\bmath
\beq
\delta B_r=-\frac{6m_ec}{e}\omega_0\frac{\lambda_L^2R}{r^3}f(r)cos\theta
\eeq
\beq
\delta B_\theta=-\frac{3 m_e c}{e}\omega_0\frac{\lambda_L^2R}{r^3}[f(r)+
\frac{r^2}{\lambda_L^2} \frac{sinh(r/\lambda_L)}{sinh(R/\lambda_L}]sin\theta
\eeq
\emath
so that the magnetic field is given by Eq. (1) except for small corrections very close to the surface.

\section{london's derivation of the London moment}
London derived the same results discussed in the previous section through a slightly different route\cite{londonbook}. He started from
the London equation
\beq
\vec{\nabla}\times\vec{v}=-\frac{e}{m_ec}\vec{B} .
\eeq
A perfect conductor of course obeys London's equation. This can be seen as follows: in the presence of an electric  field $\vec{E}$  the equation of motion for an electron in the perfect conductor is (first London equation) 
\beq
\frac{\partial \vec{v}}{\partial t}=\frac{e}{m_e}\vec{E} .
\eeq 
Taking the curl on both sides and using Faraday's law,
\beq
\frac{\partial}{\partial t} \vec{\nabla} \times \vec{v}=-\frac{e}{m_e c}\frac{\partial }{\partial t} \vec{B} .
\eeq 
Integrating and using that initially $\vec{v}=0$ and $\vec{B}=0$, Eq. (16) follows.

This derivation, which is found in essentially all superconductivity textbooks, is actually incorrect for two reasons: (i) the correct equation of motion should
use the total time derivative rather than the partial time derivative in Eq. (17), and (ii) the right-hand side of Eq. (17) should include the Lorentz force on the electron
due to a magnetic field $\vec{B}$.

The correct derivation  of Eq. (16) for a perfect conductor is as follows\cite{londonbook}: we start from the equation of motion  
\beq
\frac{d\vec{v}}{dt}=\frac{e}{m_e}\vec{E}+\frac{e}{m_ec}\vec{v}\times\vec{B} .
\eeq 
Using the relation between total and partial time  derivatives
\beq
\frac{d\vec v}{dt}=\frac{\partial \vec{v}}{\partial t}+\vec{\nabla}(\frac{v^2}{2})-\vec{v}\times (\vec{\nabla}\times\vec{v})
\eeq
and taking the curl on both sides of Eq. (19) yields
\bmath
\beq
\frac{\partial \vec{w}}{\partial t}= \vec{\nabla}\times   (\vec{v} \times\vec{w})
\eeq
with
\beq
\vec{w}=\vec{\nabla}\times\vec{v}+\frac{e}{m_e c}\vec{B}  .
\eeq
\emath
Before the perfect conductor is set into rotation, $\vec{w}=0$ since both velocity and magnetic fields are zero. According
to Eq. (21a), if $\vec{w}=0$ initially it remains equal to zero at all times. $\vec{w}=0$ is equivalent to
London's equation (16). 

Assuming that the superfluid electrons in the bulk rotate with the same angular velocity of the body
\beq
\vec{v}=\vec{\omega}_0\times\vec{r}
\eeq
replacement in Eq. (16) immediately yields
\beq
2\vec{\omega}_0=-\frac{e}{m_ec}\vec{B}
\eeq
which is the same as Eq. (1). To obtain the full solution for the current and magnetic field distribution we use Eq. (9) for the current
together with Eq. (16) and Maxwell's equation (10a) and the boundary conditions for the magnetic field at the surface of the
sphere, from which the same solution as obtained by Becker et al results, as expected.

The fundamental difference between the London and Becker et al's approaches is that London's equation (16) applies to a superconductor 
independent of its history. Thus London's derivation implies that if a rotating normal metal is cooled into the superconducting state,
the resulting velocity field and magnetic field will be given by the expressions derived by Becker et al and reviewed in Sect II. 
Instead, Becker et al would predict that a rotating metal that becomes suddenly a perfect conductor would develop no current,
since the suddenly free electrons would continue rotating at the same speed as the body because of inertia.

\section{inconsistency of London-BCS theory with measurements of the London moment}

Consider  for simplicity a superconductor where the transport of electric current occurs predominantly  in a single  band, all other bands are either completely full
or completely empty. BCS-London theory assumes that London's equation
\beq
\vec{\nabla}\times\vec{v}=-\frac{q}{m_q c}\vec{B}
\eeq
holds, where $\vec{v}$ is the superfluid velocity and $q$ and $m_q$ the charge and mass of the superfluid charge carriers in the band. The electric current density  is given by
\beq
\vec{j}=qn_s\vec{v}
\eeq
with $n_s$ the density of charge carriers. From Eqs. (24) and (25)
\beq
\vec{\nabla}\times\vec{j}=-\frac{n_sq^2}{m_q c}\vec{B}  .
\eeq
Applying the curl operator to both sides of Maxwell's equation
\beq
\vec{\nabla}\times\vec{B}=\frac{4\pi}{c}\vec{j}
\eeq
and replacing the right-hand side by Eq. (26) yields
\beq
\nabla^2\vec{B}=\frac{4\pi n_s q^2}{m_q c^2}\vec{B}\equiv \frac{1}{\lambda_L^2}\vec{B}
\eeq
which describes the exponential decay of a magnetic field in a superconductor: the magnetic field decays to zero over a distance from the surface given by the 
London penetration depth
\beq
\frac{1}{\lambda_L^2}=\frac{4\pi n_s q^2}{m_q c^2}  .
\eeq

Assume the energy band under consideration is nearly full. In the normal state, transport can be formally described using either electrons
or holes\cite{am}. However because the effective mass for electrons is negative when the band is more than half-full it is customary to describe the transport
in this case as being carried by holes, with positive charge $q>0$  and positive effective mass $m_q>0$. The simple expression for the current Eq. (25) is correct for an almost full band
only if $n_s$ denotes the density of $holes$, not of electrons\cite{am}. Furthermore, when a band is almost full, the London penetration depth
is found to diverge, which is consistent with the number of superfluid charge carriers going to zero in Eq. (29), i.e. with $n_s$ describing the density
of holes rather than of electrons in the band. In the same way, the number of carriers that enters the normal state conductivity in the
normal state within Drude theory is the number of holes rather than the number of electrons when the band is almost full, using
the fact that a full band carries no current (nor supercurrent). 
Note that the London penetration depth is independent of the $sign$ of the charge carriers.

For example, consider an attractive Hubbard model with Hamiltonian
\beq
H=-t\sum_{i,\sigma}[c_{i\sigma}^\dagger c_{i\sigma} + h.c.]+U\sum_i n_{i\uparrow}n_{i\downarrow}
\eeq
with $U<0$ as a simple model to describe conventional superconductivity in a real system. Straightforward application of BCS theory\cite{tinkham} yields
for the London Kernel describing the diamagnetic response of the system, at zero temperature\cite{londonhm}
\beq
K=\frac{1}{\lambda_L^2}=\frac{8\pi e^2 t}{\hbar^2c^2 a_\delta} \frac{1}{N}\sum_k cos k_\delta [1-\frac{\epsilon_k-\mu}{E_k}]
\eeq
with $a_\delta$ the lattice spacing in the $\delta$ direction and $E_k=\sqrt{(\epsilon_k-\mu)^2+\Delta^2}$ the BCS quasiparticle
energy, with $\Delta$ the energy gap. This can also be written as
\beq
K=\frac{1}{\lambda_L^2}=\frac{8\pi e^2 t}{\hbar^2c^2 a_\delta} \frac{1}{N}\sum_k cos k_\delta n_k
\eeq
with $n_k$ the occupation of state $k$. When the band is almost full we have simply
\beq
\frac{1}{N} \sum_k  cos(k_\delta) n_k  \sim n_h
\eeq
with $n_h$ the number of $holes$ per site. The effective mass for holes near the top of the band in this model, with energy dispersion
relation
\beq
\epsilon_k=-2\sum_\delta cos(k_\delta a_\delta)
\eeq
is
\beq
\frac{1}{m_q}=\frac{1}{\hbar^2}\frac{\partial ^2 \epsilon_k}{\partial k_\delta ^2} )_{k=0}=\frac{2t a_\delta^2}{\hbar^2}
\eeq
Hence from Eqs. (32), (33) and (35) and assuming an isotropic model so that the density of superfluid holes (of positive
charge $q=-e$ per unit volume) is
\beq
n_s=\frac{n_h}{a_\delta^3}
\eeq
we find
\beq
\frac{1}{\lambda_L^2} =\frac{4\pi n_s q^2}{m_qc^2}    .
\eeq
The result Eq. (37) is identical to Eq. (29) obtained through the electrodynamic equations. Therefore we conclude that
Eq. (24) applies to this model for the superfluid hole carriers, with $q>0$ and effective mass $m_q$ given by Eq. (35).
Following then London's derivation, for a rotating superconductor the velocity of the hole carriers in the interior of the body
is identical to the ionic velocity, i.e. $\vec{v}=\vec{\omega}_0\times \vec{r}$ and we obtain
\beq
\vec{B}=-\frac{2 m_q c}{q} \vec{\omega_0}
\eeq
which is different from the measured value Eq. (1) both in sign and in magnitude. In particular, Eq. (38) predicts that the
magnetic field of a rotating superconductor described by this model is $antiparallel$ to the angular velocity rather than
parallel as observed, since $q>0$.

The reader may argue that an attractive Hubbard model with an almost full band is not an accurate description of any real
superconductor. We argue that in fact superconductors are almost always found to have positive Hall coefficients in the
normal state\cite{chapnik},  or at least to have one band with hole carriers in a multi-band situation\cite{hall2}, 
and that conduction of one particular band often dominates the transport, so that an approximate description 
with a model with a single almost full band is appropriate in many cases. The use of an instantaneous attractive interaction instead
of a retarded electron-phonon one is a simplifying approximation that may be quantitatively appropriate only for high frequency
phonons but in any event does not qualitatively affect the results, as can be seen for example in the calculation of
London penetration depth in a lattice electron-phonon model using Eliashberg theory as in Ref.\cite{londonep}. Whether the
superconductor is $s-$ or $d-$wave as believed to be for certain unconventional superconductors would not change
our results qualitatively either, as can be seen from refs. \cite{londonmw1,londonmw2}.

\section{the unexplained and counterintuitive features of the London moment}
In this section we summarize the unexplained features of the London moment within the conventional BCS-London theory of
superconductivity.

\subsection{The sign}
The measured direction of the magnetic field generated by rotation is always $parallel$, never $antiparallel$, to the angular 
velocity\cite{lm1,lm2,lm3,lm4,lm5,lm6,lm7,lm8}.
According to our discussion in the previous section, if the carriers in the normal state have dominant hole character the resulting
magnetic field should be antiparallel to the angular velocity according to conventional theory. In other words, experiments tell us that
superfluid carriers are always negatively charged electrons, while normal state transport experiments in superconductors most often
yield positive Hall coefficient indicating that normal state carriers are hole-like, positive. BCS theory does not predict a change in the
character of electronic states from hole-like to electron-like  when a system goes superconducting, hence the observations are
inconsistent with BCS theory.

\subsection{The magnitude}
The magnitude of the measured magnetic field is given by Eq. (1) with $m_e$ the $bare$ electron mass. This is inconsistent with the prediction of 
BCS theory as reviewed in the previous section, that the mass that enters in the expression for the London penetration depth
Eq (37) is the effective rather than the bare mass. The same mass that enters in the expression for the London penetration depth should enter
in Eq. (24) from which the London field Eq. (1) directly follows upon applying the curl. Note that the effective mass in heavy fermion systems can
be up to 3 orders of magnitude larger than the bare electron mass, and the London moment measured in heavy fermion systems is
no different from that measured in conventional superconductors\cite{lm8}.

We can see the problem most directly in the  simple derivation Eqs. (3) to (8). Eq. (3) is valid for an electron in free space but not for an electron
in the conduction band of a solid. In a solid, semiclassical transport theory\cite{am} tells us that the equation  of motion is, instead of Eq. (3)
\beq
\hbar\dot{\vec{k}}=eE
\eeq
and
\beq
v=\frac{1}{\hbar} \frac{\partial \epsilon}{\partial k}
\eeq
from which
\bmath
\beq
 \frac{\partial v}{\partial t}=\frac{1}{\hbar^2} \frac{\partial ^2 \epsilon_k}{\partial k^2}eE =  \frac{e}{m^*}E
 \eeq
 \beq
 \frac{1}{m^*}=\frac{1}{\hbar^2} \frac{\partial ^2 \epsilon_k}{\partial k^2}
 \eeq
 \emath
 so that Eqs. (4) to (8) follow with $m^*$ replacing $m_e$, and the result instead of Eq. (8) is
 \beq
B=-\frac{2m^* c }{e} \omega_0
\eeq
in contradiction with experiment, since the effective mass is never  identical  to the bare mass. In other words, the response of an electron in the conduction
band of a solid, whether the electric field is an externally applied field derived from an applied voltage difference or an electric field arising from
Faraday's law and a changing magnetic field generated by the ionic current of the rotating body, is described in terms of the effective mass and
not the bare mass.

Thus, the  London moment measurements tell us that the conduction electrons in the solid become completely `free' of interactions with the ionic lattice in the 
superconducting state. This point of view was prevalent in the early days of superconductivity. In particular, Becker et al\cite{becker} start their paper by
stating ``Man pflegt den supraleitenden Zustand dadurch zu deuten, dass man den Leitungselektronen relativ zu den Gitterionen eine
unendlich grosse Beweglichkeit zuschreibt'' (``It is customary to  construe  the superconducting state by attributing  an infinite mobility to the electrons relative to the
lattice of ions''). 
It was pointed out for example by Rudnitzkij\cite{rud} that the London moment experimental measurement proves that
the electron becomes free in the superconducting state. BCS theory does not describe any change in the character of the electronic 
states other than the pairing correlations, and the electric current in the normal state is carried by carriers with effective mass $m^*$ rather than bare mass $m_e$,
so BCS theory is in disagreement with experiment. Furthermore, BCS theory attributes the pairing of the carriers to the electron-phonon interaction, which is 
completely inconsistent with the evidence that superfluid carriers become free of interactions with the ionic lattice.

\subsection{The slowing down of electrons near the surface}
In a rotating normal metal the electrons move together with the ions, there is no current and no magnetic field. When the metal is cooled into
the superconducting state, electrons near the surface spontaneously slow down to give rise to the electric current 
\beq
\vec{j}=3en_s \vec{\omega}_0\times \vec{r} \frac{\lambda_L^2 R}{r^3} f(r)
\eeq
(with $f(r)$ given by Eq. (11b)) that gives rise to the magnetic field Eq. (1). What is the physical mechanism that causes the electrons to slow down? As recounted in the introduction, London 
predicted this confidently even though he viewed it as ``quite absurd from the viewpoint of the perfect conductor concept'' and offered no 
physical explanation for why this would occur.

There is another associated puzzle. The angular momentum of the entire system cannot change upon cooling. The electrons near
the surface move slower and consequently carry smaller angular momentum than in the normal state. Hence we have to conclude that
the angular velocity of the body slightly $increases$ in the transition from the normal to the
superconducting state. What is the physical explanation for this phenomenon?

BCS theory does not provide us with an explanation of these puzzles, it merely agrees with London theory in that this is the unique state
of the rotating superconductor. But how does the rotating normal metal  get there? BCS theory does not offer any clues.

If  it is ever experimentally observed that a rotating normal metal cooled into the superconducting state does $not$  develop a magnetic field, one could 
not say that this would falsify BCS theory, one could simply say that the system has remained in an infinitely long-lived metastable state which is not the lowest energy state according to BCS theory. By the same token it can be said that BCS theory does not predict nor explain the development of the London moment.

\subsection{The radial electric field}

In the rotating superconductor the electrons experience a Lorentz force due to the magnetic field Eq. (1):
\beq
F_B=\frac{e}{c} v B
\eeq
For electrons  moving at the same speed as the ions this force is, using Eq. (1)
\beq
F_B=2m_e \omega_0^2 r
\eeq
Here and in what follows we use $r$ to denote distance to the rotation axis rather than radius in spherical coordinates.
This force points $inward$ in the direction of the rotation axis. To sustain the rotational motion, a centripetal force is required
\beq
F_c=m_e\frac{v^2}{r}=m_e \omega_0^2 r
\eeq which is only half of Eq. (45). Thus, the Lorentz force will pull electrons slightly $inward$, until a radial electric field is created to
balance the radial forces. The resulting electric field is  

\beq
E_r=\frac{m_e}{e}\omega_0^2 r
\eeq
giving rise  to an outward force on electrons
\beq
F_E=eE_r=m_e\omega_0^2r
\eeq
thus achieving  radial force balance:
\beq
F_B=F_c+F_E
\eeq
The situation is depicted schematically in Fig. 2.

  \begin{figure}
\resizebox{8.5cm}{!}{\includegraphics[width=7cm]{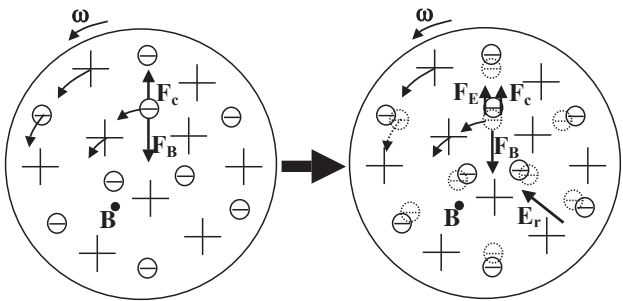}}
  \caption{The electrons (full lines) move slightly inward (dotted lines)  pulled in by the magnetic Lorentz force
  $\vec{F}_{\vec{B}}$ , creating an inward-pointing electric field
 $\vec{E}_r$  that balances the radial forces. }
\end{figure} 

Assuming the electrons become more `free' from the ions in the superconducting state, one
would expect that they  move $out$ rather than $in$  in the rotating superconductor due to the centrifugal force.
This was the expectation of Becker et al\cite{becker}, they
stated: ``Wir beschr\"anken unsere Betrachtungen von vornherein auf die Rotationsbewegung
der Elektronen, vernachl\"assigen also die dureh die Zentrifugalkraft
bedingte negative Aufladung der Kugeloberflache'', meaning 
``We restrict our considerations from the beginning to the rotational motion of the electrons, neglecting the negative charging of the surface 
caused by the centrifugal force''. In references \cite{rys} and \cite{gaw} it was also concluded that negative charges would move $out$ because
of an incorrect analysis of the radial forces involved\cite{spinc}, and the authors presumably did not double-check their result relying on the
same physical intuition  as Becker et al.

That physical intuition is in fact flawed for the case of a perfect conductor. A perfect conductor starting from rest would indeed develop a magnetic field as
the electrons near the surface lag behind the motion of the ions, and the Lorentz force due to this magnetic field would pull the electrons slightly
inward. This physical effect is well known in plasma physics under the name `theta-pinch'. However, here we want to focus on the London
scenario.

The question whether negative charges move inward or outward in a rotating superconductor has never been explored experimentally.
According to London-BCS theory, the answer has to be the same whether the superconductor is set into motion or whether the rotating
normal metal is cooled into the superconducting state. For the latter case it  certainly   defies physical intuition that the superfluid charges would 
spontaneously move inward, just as it defies intuition that the azimuthal motion would spontaneously slow down. Thus, we argue that this prediction
of the conventional theory involving inward radial motion as depicted in Figure 2 is likely to be incorrect.

\section{A simple way to explain the key mystery  of the London moment}

The most puzzling observation related to the London moment, ``quite absurd from the viewpoint of the perfect conductor concept'' according to 
London, is that electrons near the surface spontaneously slow down when a rotating metal is cooled and becomes superconducting.

Superconductivity cannot suspend the laws of mechanics and electromagnetism. Inertia compells bodies to continue their motion unless a force acts
to modify the motion. What is the azimuthal  force causing electrons near the surface to spontaneously slow down? There is no mechanical force nor 
electromagnetic force to do this task. BCS theory does not offer any insight into the process of slowing down of electrons near the surface.

There is however a very simple, intuitive explanation for this phenomenon: assume that a {\it radial outflow}  of electrons takes place when a metal goes superconducting.
Then, electrons from the interior that reach the surface will be moving at a slower tangential velocity than the surface, reflecting the fact that
the tangential velocity of the body is slower in the interior. This is shown schematically in Fig. 3. The slower velocity of electrons
near the surface reflects the fact that they were originally  deep in the interior of the body and following the body's rotation. As viewed
in  the framework of the rotating body, electrons moving radially outward with velocity $\vec{v}_r$ experience a Coriolis force 
$\vec{F}_c=-2m_e \vec{\omega}_0\times\vec{v}_r$ in azimuthal direction opposite to the body's motion.

As the electrons move out, they will decrease their tangential velocity if they keep their angular momentum constant. Alternatively, they could
end up with any azimuthal velocity up to the azimuthal velocity they had in the interior, in which case   the body as a whole   will slightly slow down to conserve total angular momentum. This will depend on the details of the process, but in any case the azimuthal velocity of electrons that came
from the interior will be smaller than the azimuthal velocity of the body at the surface, thus explaining the origin of the surface current for the
case when a rotating normal metal is cooled into the superconducting state..

Note that a $force$ is also needed to explain the process just outlined. However, for this process we need a $radial$ force rather than
an azimuthal force. One could imagine different sources, for example that the $centrifugal$ $force$ in the rotating frame becomes more
effective in  the superconducting versus the normal state. In fact, we believe there is another radial force at play, 
originating in {\it quantum pressure}, as discussed in the
next section.

  \begin{figure}
\resizebox{6.5cm}{!}{\includegraphics[width=7cm]{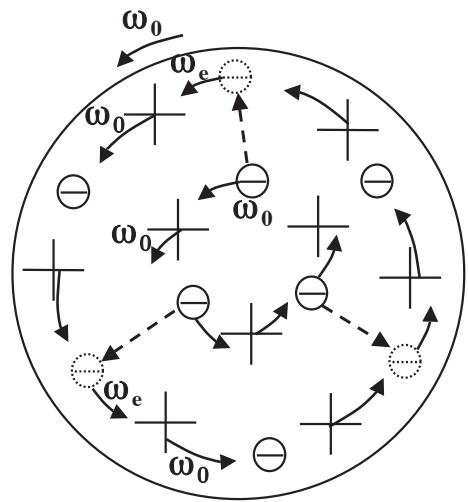}}
  \caption{Electrons that move from the interior to the surface  have smaller tangential velocity
  than the ions near the surface because they came from a region where the tangential velocity was smaller, giving rise to an electric current near
  the surface which creates a magnetic field $parallel$ to the body's angular
  momentum. }
\end{figure}

\section{The London moment within the theory of hole superconductivity}

The theory of hole superconductivity\cite{holesc} proposes that superconductivity in solids originates in the fundamental 
charge asymmetry of matter\cite{sitges},
namely that positive charge (protons) is heavier than negative charge (electrons). It predicts that superconductivity   can only occur
when the charge carriers in the normal state are holes. Thus, within this theory superconductors know very well the difference between
electrons and holes, or between negative and positive charge, in contrast to conventional BCS theory that can be formulated
with models that are electron-hole symmetric. As discussed earlier, the London moment (by being always of one sign) reveals
that real superconductors know very well the difference between positive and negative charge\cite{rotating}, in agreement with the theory of
hole superconductivity and in disagreement with BCS theory. 

In addition, several essential features of the theory of hole superconductivity explain the puzzles of rotating superconductors discussed earlier:

\subsection{Negative charge expulsion}
The theory predicts that metals expel negative charge from the interior to the surface in the transition to superconductivity\cite{chargeexp}.
The prediction follows from the microscopic Hamiltonian used in the theory\cite{hm89,dynhubexp} (dynamic Hubbard model\cite{dynhub}) and the resulting form of the
gap function\cite{hm89,imbalance}, as well as from alternative electrodynamic equations proposed within the theory\cite{electrodyn}. This is predicted to occur independent
of whether the body is rotating or not and independent of whether or not an external magnetic field is applied. In the presence of an external
magnetic field it provides a dynamical explanation for the Meissner effect\cite{meissner}, and in the presence of body rotation it provides an explanation
for the slowing down of electrons near the surface discussed in the previous section.

\subsection {Superconductivity from `undressing'}
The theory predicts that in superconductors in the normal state charge carriers are heavily `dressed' by both electron-electron interactions
and electron-ion interactions\cite{holeelec2}. Both `dressings' are largest when the electronic energy band is almost full. In particular the electron-ion
`dressing' is what changes the sign of the effective mass from its bare value (positive) to its dressed value (negative). The theory  furthermore predicts
that in the transition to superconductivity carriers `undress'  from both the electron-electron and the electron-ion interaction\cite{undr,hvar}.
The microscopic calculations performed so far give rise only to partial `undressing' of the carriers\cite{undr}, however physical considerations
led us to conclude that complete undressing occurs in the transition to superconductivity and the superfluid carriers behave as free
electrons\cite{hvar}. This is consistent with the observation that the bare mass of the electron enters in the expression for the London moment
rather than the effective mass.

\subsection {Orbit expansion}
The theory predicts that electronic orbits expand in the transition to superconductivity, from microscopic radius
$k_F^{-1}$ ($k_F=$Fermi wavevector) to mesoscopic radius $2\lambda_L$\cite{sm}. In particular, this describes the increase  in the
diamagnetic susceptibility from Landau's value for normal metals to $-1/4\pi$ for superconductors\cite{meissner}. 

Clearly, if superconducting electrons reside in orbits of radius several hundred  $\AA$`s ($2\lambda_L$) it is to be expected that they don't `see' the microscopic
ionic structure which changes over a scale of a few $\AA$ and instead see an average continuous positive charge distribution. 
As a consequence they are `undressed' from the electron-ion interaction and behave as free electrons, which is reflected in the fact that
their bare rather than their effective mass enters into the London field expression Eq. (1).

\subsection {Holes becoming electrons}

The theory of hole superconductivity predicts that a redistribution of occupation of single electron states occurs in the transition
to superconductivity\cite{eh3}. Namely, that the holes that reside on top of the electronic energy band migrate to the bottom of the 
electronic energy band. This gives rise to an enlargement of the wavelength associated with the charge carriers and
thus explains the `undressing' from the electron-ion interaction. Furthermore, the holes at the bottom of the band have now dispersion
relation of opposite curvature as at the top of the band, hence they respond as electrons rather than as holes. In particular,
their Hall coefficient, if it could be measured, would be negative. However, the {\it number of charge carriers} has not changed, it is
still the number of $holes$ in the band, going to zero as the band becomes full. This explains the puzzle of why the London moment
appears to be generated by $electrons$ with negative charge, yet the $number$ of charge carriers $n_s$ as reflected
in the London penetration depth is the number of holes in the band.

\subsection { Radial motion and existence of spin current}

Assume that in the absence of body rotation electrons are moving in large orbits centered at the rotation axis, with azimuthal speed $v_e$. The centripetal acceleration for this
motion is provided by the outward pointing electric field resulting from charge expulsion discussed in A. When the body is set into rotation, the azimuthal speed
will change by $\Delta v_e=\omega_0 r$ in the interior. The centripetal acceleration will change by
\beq
\Delta(\frac{m_e v_e^2}{r})=\frac{2 m_e v}{r}\Delta v_e=2m_e\omega_0 v_e=\frac{e}{c}v_e B
\eeq
In the last equality we have used the expression for the London field Eq. (1).  Thus, we obtain that the increase  in centripetal force when the body rotates is
$exactly$ $canceled$ by the inward Lorentz force due to the magnetic field in the rotating superconductor. In other words, no radial redistribution occurs
when a superconductor at rest is put into rotation. When a rotating normal metal is cooled into the superconducting state of course radial redistribution
occurs because electrons are expelled from the interior to the surface. In any event, the unphysical radial inward motion predicted by the
conventional viewpoint (Fig. 2) does not take place in this scenario.

If in the superconductor electrons are rotating in large orbits in the absence of applied fields and rotation, this has to occur in the absence of charge currents.
Thus, we conclude that for every electron orbiting to the right there has to be another electron orbiting to the left. This is consistent with absence of a charge current
but it allows for the existence of a {\it spin current}, where electrons of opposite spin orbit in opposite directions. This is what happens in the superconducting
state according to the theory of hole superconductivity\cite{spinc,sm,electrospin}. Thus, within this theory there is no puzzle associated with inward radial motion of electrons in the
rotating superconductor.

\subsection{Kinetic energy lowering}

When an electron expands its orbit its kinetic energy decreases, since quantum kinetic energy is given by 
$\hbar^2/(2m_e r^2)$. The theory of hole superconductivity says that superconductivity is associated with lowering of
kinetic energy\cite{kinenergy}, in contrast with BCS theory that says that kinetic energy is increased in the transition to the superconducting state.

Kinetic energy lowering is associated with quantum pressure\cite{emf}, the tendency of a quantum particle to $expand$ its wave function
to lower its kinetic energy. It can also be understood from the uncertainty principle. This quantum pressure is associated with
a $radial$ $force$ ($=$ - (change in kinetic energy) with radial expansion). Thus we propose that this is the force behind
the predicted negative charge expulsion, and it is the force that ultimately explains why electrons near the surface
`slow down' when the rotating normal metal becomes superconducting through the mechanism discussed in Sect. VI.

   \begin{figure}
\resizebox{8.5cm}{!}{\includegraphics[width=7cm]{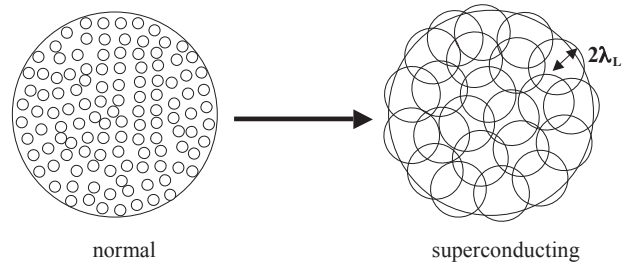}}
  \caption{ In the normal state (left), electrons reside in non-overlapping orbits of radius $k_F^{-1}$. The orbits expand to radius $2\lambda_L$ in the
  superconducting state (right). The electros at the surface have the center of their orbit a distance $2\lambda_L$ from the surface, i.e. at
  radius $R-2\lambda_L$, and thus have a smaller tangential velocity.}
\end{figure}

 \section{$2\lambda_L$ orbits and slowing down near the surface}
 Within the theory of hole superconductivity in the transition from the normal to the superconducting
 state, electronic orbits expand from radius $k_F^{-1}$ to radius $2\lambda_L$. This is accompanied by outward motion of charge, and if there is a magnetic
 field present it is expelled as the orbits expand, giving rise to the Meissner effect.
 
 We can similarly understand the slowing down of electrons near the surface when the rotating normal metal is cooled into the superconducting state. For the
 sphere, the slowing down is given by
 \beq
 \vec{v}_s-\vec{v}_0=-\frac{3\lambda_L}{R}\vec{\omega}\times\vec{r}
 \eeq
 and for a cylinder by\cite{laue}
  \beq
 \vec{v}_s-\vec{v}_0=-\frac{2\lambda_L}{R}\vec{\omega}\times\vec{r}
 \eeq
 Just like for the Meissner effect, it is cylindrical rather than spherical geometry that allows for the simplest understanding. This illustrated by the fact that the critical
 magnetic field for a cylindrical geometry is $H_c$, while it is $(2/3) H_c$ for a sphere due to demagnetization.
 
 The orbit expansion is shown schematically in Fig. 4. The electrons at the surface have their center of motion at distance $2\lambda_L$ from the surface. Thus, 
 their rotation velocity at the surface corresponds to the tangential velocity of rotation at radius $R-2\lambda_L$, i.e.
 \beq
 v_s=\omega(R-2\lambda_L)
 \eeq
 in agreement with Eq. (52). In other words, electrons at the surface move slower because they were originally at radius $R-2\lambda_L$, with
 slightly slower tangential velocity.

\section{Summary and discussion}

Perhaps the reason that the London moment is not discussed in standard superconductivity textbooks is because of the difficulty
in explaining the observations with the conventional theory of superconductivity discussed in those textbooks, as argued in this paper.

Let us summarize the main points made in this paper:
\newline
(1) Superconductors know the difference between positive and negative charge, unlike normal metals, as demonstrated by universal
sign of the London moment. We argue that this is a  strong point against electron-hole symmetric theories like conventional 
London-BCS theory and in favor of the theory of hole superconductivity that has as its foundation electron-hole asymmetry.
\newline
(2) The London moment shows that superfluid electrons behave as bare electrons, free of interactions with the ionic lattice.
This is inconsistent with conventional London-BCS theory and particularly with the assumed electron-phonon origin of pairing
for conventional superconductors. It is consistent with the theory of hole superconductivity that predicts undressing of carriers
as the system enters the superconducting state. 

We should mention that there is other other experimental evidence for this undressing: (i) the change in sign of the Hall coefficient
from positive to negative as a metal is cooled into the superconducting state\cite{hallch}, and (ii) for high $T_c$ cuprates the observation
in photoemission experiments that the quasiparticle weight strongly increases upon entering the superconducting state\cite{z1,undr},
an effect not predicted by BCS theory.
\newline
(3) The key puzzle of the London moment experiment, that electrons near the surface `slow down' when a rotating normal metal
becomes superconducting, and as a consequence  the body as a whole speeds up to conserve angular momentum, 
is unexplained within London-BCS theory. The fact that electrons slow down near the surface is simply explained by the hypothesis that electrons move from the interior to the surface of the sample
in the transition to superconductivity, in which case it is not necessary for the body to speed up to conserve angular momentum. This is predicted by the theory of hole superconductivity.
\newline
(4) Quantitatively, the magnitude of the slowing down near the surface is consistent with electrons at $r=R$ coming from an interior position at $r=R-2\lambda_L$. This
is in turn consistent with the prediction of the theory of hole superconductivity that superfluid electrons had their orbits enlarged from a microscopic scale 
to orbits of radius $2\lambda_L$.

 We suggest that the arguments presented in this paper should lead to questioning the validity of conventional London-BCS theory to
 explain superconductivity\cite{validity}, given that the London moment is a universal effect in superconductors just as the Meissner effect.
 Alternatives to conventional London-BCS theory that are consistent with the London moment observations should be proposed and
 critically examined. The theory of hole superconductivity is suggested as one possibility.

\acknowledgements
The author is grateful to  O. Shpyrko   for translating Ref. \cite{rud} from Russian to English.

 \end{document}